\documentclass[aj,twocolappendix,appendixfloats,numberedappendix]{emulateapj}

\usepackage{graphicx}
\usepackage{epstopdf}
\usepackage{rotating}
\usepackage{url}
\usepackage{appendix}

\begin{document}

\title{X-ray Observations of Bow Shocks around Runaway O Stars.\\ The
  case of $\zeta$\,Oph and BD$+$43$^{\circ}$3654.}
\author{J.A.\,Toal\'{a}$^{1}$\footnotemark[$\ast$],
  L.M.\,Oskinova$^{2}$, A.\,Gonz\'{a}lez-Gal\'{a}n$^{2}$,
  M.A.\,Guerrero$^{1}$, R.\,Ignace$^{3}$, and M.\,Pohl$^{2,4}$}

\affil{$^{1}$Instituto de Astrof\'{i}sica de Andaluc\'{i}a, IAA-CSIC,
  Glorieta de la Astronom\'{i}a s/n, 18008 Granada,
  Spain; toala@iaa.es\\ 
$^{2}$Institute for Physics and Astronomy, University of
  Potsdam, 14476 Potsdam, Germany\\ 
$^{3}$Department of Physics and
  Astronomy, East Tennessee State University, Johnson City, TN 37614,
  USA\\ 
$^{4}$DESY, Platanenallee 6, 15738 Zeuthen, Germany
}

\footnotetext[$\ast$]{Now at: Institute of Astronomy and Astrophysics,
  Academia Sinica (ASIAA), Taipei 10617, Taiwan.}

\begin{abstract}
  Non-thermal radiation has been predicted within bow shocks around
  runaway stars by recent theoretical works. We present X-ray
  observations towards the runaway stars $\zeta$\,Oph ({\it Chandra}
  and {\it Suzaku}) and BD+43$^{\circ}$3654 ({\it XMM-Newton}) to
  search for the presence of non-thermal X-ray emission. We found no
  evidence of non-thermal emission spatially coincident with the bow
  shocks, nonetheless, diffuse emission is detected in the vicinity of
  $\zeta$\,Oph. After a careful analysis of its spectral
  characteristics we conclude that this emission has a thermal nature
  with a plasma temperature of $T\approx2\times10^{6}$~K. The cometary
  shape of this emission seems to be in line with recent predictions
  of radiation-hydrodynamic models of runaway stars. The case of
  BD+43$^{\circ}$3654 is puzzling as non-thermal emission has been
  reported in a previous work for this source.
\end{abstract}


\section{Introduction} 
\label{sec:intro}

Runaway stars are thought to be ejected from their formation nursery
with high velocities
\citep[$v_*\gtrsim30$\,km\,s$^{-1}$;][]{Gies1986,Tetzlaff2011}. The
origin of these high velocities is still a matter of debate. Some
possibilities include the effects of close interactions between binary
systems in a cluster \citep[e.g.,][]{Hoogerwerf2000}, strong
gravitational interactions between single and binary systems
\citep[e.g.,][]{Fujii2011}, or kicks arising from a supernova
explosion of a binary companion \citep[e.g.,][]{Blaauw1961}.

Runaway massive ($M >10M_{\odot}$) stars moving supersonically through
the interstellar medium (ISM) produce large-scale bow shocks. The gas
and dust compressed in bow shocks is heated and ionized by the intense
stellar radiation making these large scale ISM structures observable
in infrared (IR) and in optical (e.g. H$\alpha$) emission
\citep[e.g.,][]{vanBuren1988}. Indeed, many stellar bow shocks have
been detected in optical and IR wavelengths
\citep[e.g.,][]{vanBuren1995,Kaper1997,NoriegaCrespo1997,Kobulnicky2010,
  Peri2012}.  Nevertheless, there are certain physical conditions for
which a stellar bow shock may not form. For example, if the star is
moving with sub-sonic velocities in a too tenuous, hot ambient medium
or if it has a weak wind or a high space velocity
\citep[e.g.,][]{Comeron1998,Huthoff2002}.

Bow shocks around massive stars are also detected at radio
wavelengths. \cite{Benaglia2010} reported, for the first time, radio
emission from the bow shock around a massive runaway star
(BD+43$^\circ$3654). Their Very Large Array (VLA) observations
provided a stark evidence that non-thermal radio emission is spatially
coincident with the bow shock observed in infrared
emission. \citet{Benaglia2010} argued that this non-thermal emission
should arise from the cooling of energetic electrons by syncrotron
emission. The electrons that produce this non-thermal radio emission
could upscatter photons from the stellar and dust photon fields via
the inverse Compton process, leading to high-energy emission. In
particular, inverse Compton scattering into the X-ray band requires
very low-energy electrons with Lorentz factors of the order of 100.

Since this discovery, a number of theoretical works have been
presented to address the production of non-thermal emission at the
position of the bow shock around runaway stars \citep[see][and
references therein]{delValle2015}. \cite{delValle2012} presented
analytical calculations with applications to the closest runaway
massive O-type star, $\zeta$ Oph, and concluded that non-thermal X-ray
and $\gamma$-ray emission from its bow shock should be
detectable. This work was farther extended in \citet{delValle2014}
where the model spectral energy distribution over the broad range of
energy was presented.

\begin{figure*}
\begin{center}
\includegraphics[angle=0,height=0.5\linewidth]{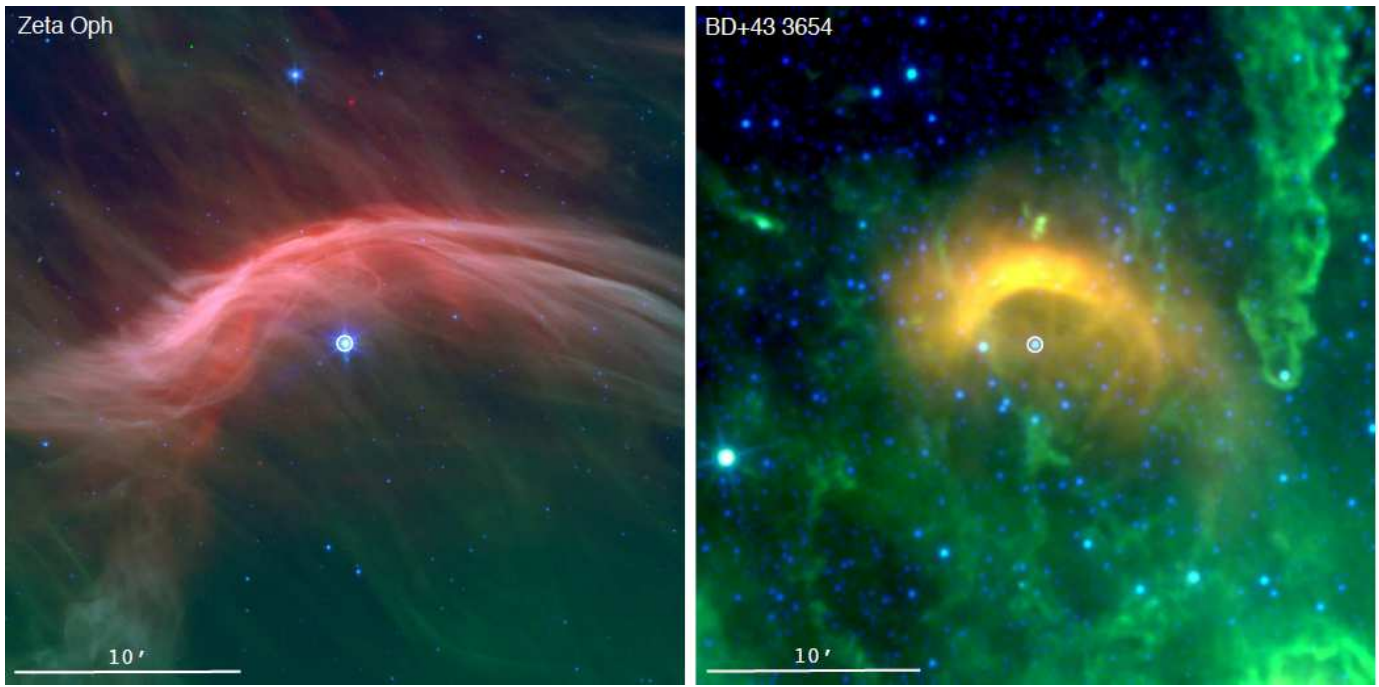}
\caption{Color-composite mid-IR images of $\zeta$\,Oph (left panel)
  and BD$+$43$^{\circ}$3654 (right panel). For $\zeta$\,Oph red,
  green, and blue correspond to {\it Spitzer} MIPS 24~$\mu$m, IRAC
  8~$\mu$m, and IRAC 4.5~$\mu$m, respectively. In the case of
  BD$+$43$^{\circ}$3654 red, green, and blue correspond to {\it WISE}
  22, 12, and 4.6~$\mu$m, respectively. The circular aperture in both
  panels show the position of the central stars. North is up and east
  is left.}
\end{center}
\label{fig:WISE}
\end{figure*}

The predictions of del Valle \& Romero's model were observationally
tested by \citet{Schulz2014}. Using the analysis of data accumulated
during 57 months by the Fermi $\gamma$-ray Space Telescope, the first
systematic search of $\gamma$-ray emission from 27 bow shocks around
runaway stars was performed.  No positive detections were obtained. It
was demonstrated that for the case of $\zeta$\,Oph the upper limit on
its $\gamma$-ray emission is 5 times below that predicted by
\citet{delValle2012}. At the X-ray wavelengths, \citet{Terada2012}
presented {\it Suzaku} observations of BD$+$43$^{\circ}$3654 and did
not detect non-thermal X-ray emission associated to its bow
shock. Only one marginal detection of non-thermal X-ray emission from
a bow shock around a runaway star has been reported to date. This
detection was claimed by \cite{LopezSantiago2012} for AE Aurigae (HIP
24575) using {\it XMM-Newton} observations. Unfortunately, the
data did not allow to discriminate between non-thermal and thermal
emission.

Besides non-thermal radiation, X-rays from bow shocks originating in
thermal plasma can be expected. A number of numerical simulations have
shown that the stellar wind-ISM interaction resulting from (slow and
fast) moving stars produce instabilities that mix material between the
adiabatically shocked wind and the photoionized gas at the wake of the
bow shock
\citep[e.g.,][]{BDE1995a,BDE1995b,Arthur2006,Mackey2015,Meyer2015};
this creates a mixing region capable of producing plasma temperatures
of $\sim10^{6}$~K. In particular, \citet[][]{Mackey2015} presented
radiation-hydrodynamic simulations on the formation of bow shocks
around massive O-type stars and showed that these instabilities are
capable of produce diffuse X-ray emission at the wake\footnote{Note,
  however, that the simulations presented by \citet{Mackey2015} are
  tailoled to runaway stars with velocities
  $v_{\star}$=4--16~km~s$^{-1}$.}.

In this paper we present {\it Chandra}, {\it Suzaku} and {\it
  XMM-Newton} observations towards the runaway O stars $\zeta$\,Oph
and BD+43$^{\circ}$3654 to explore the existence of extended X-ray
emission associated to their bow shocks and its nature. Both of these
runaway stars display extended bow shocks seen in mid-infrared images
(see Figure~1) and they are relatively close and suffer only modest
extinction, allowing to probe soft X-ray emission. This makes them the
best candidates to test the predictions from theory.

The wind parameters of $\zeta$ Oph (O9.2IV) were derived by
\citet{Marcolino2009} from modeling its optical and UV spectra:
$\dot{M}\approx 1.6\times 10^{-9}\,M_\odot$\,yr$^{-1}$ and terminal
wind velocity $v_\infty\approx 1500$\,km\,s$^{-1}$, while from the
analysis of its bow shock \citet{Gvar2012} find an order of magnitude
higher mass-loss rate $\dot{M}\approx 2\times
10^{-8}\,M_\odot$\,yr$^{-1}$. This discrepancy could be explained if
the bulk of $\zeta$ Oph wind is in a hot phase \citep{Hue2012}. In
this case, the wind kinetic power is $E_{\rm wind}\approx 1.4\times
10^{34}$\,erg\,s$^{-1}$.

From the analysis of the IR image of the bow shock around
BD+43$^{\circ}$3654 (O4If), \citet{Kobulnicky2010} found a very large
mass-loss rate for this star $\dot{M} \sim 2\times
10^{-4}\,M_\odot$\,yr$^{-1}$. However they pointed out that this value
is uncertain because of poorly known ISM density around this
object. The mass-loss rate of an O-type star with the same spectral
type, $\zeta$\,Pup (O4If(n)) is $\dot{M}\approx 2.5\times
10^{-6}\,M_\odot$\,yr$^{-1}$ and wind velocity $v_\infty\approx
2250$\,km\,s$^{-1}$ \citep{Oskinova2007,Surlan2013}. Adopting these
parameters, results in a wind power of $E_{\rm wind}\approx 4\times
10^{36}$\,erg\,s$^{-1}$.

The structure of the paper is as follows. In Section~2 we describe the
X-ray observations. Section~3 gives a description of the results and
spectral analysis. We discuss our findings and present our conclusions
in Sections~4 and 5, respectively.

\section{Observations}

\begin{figure*}
\begin{center}
\includegraphics[angle=0,width=0.5\linewidth]{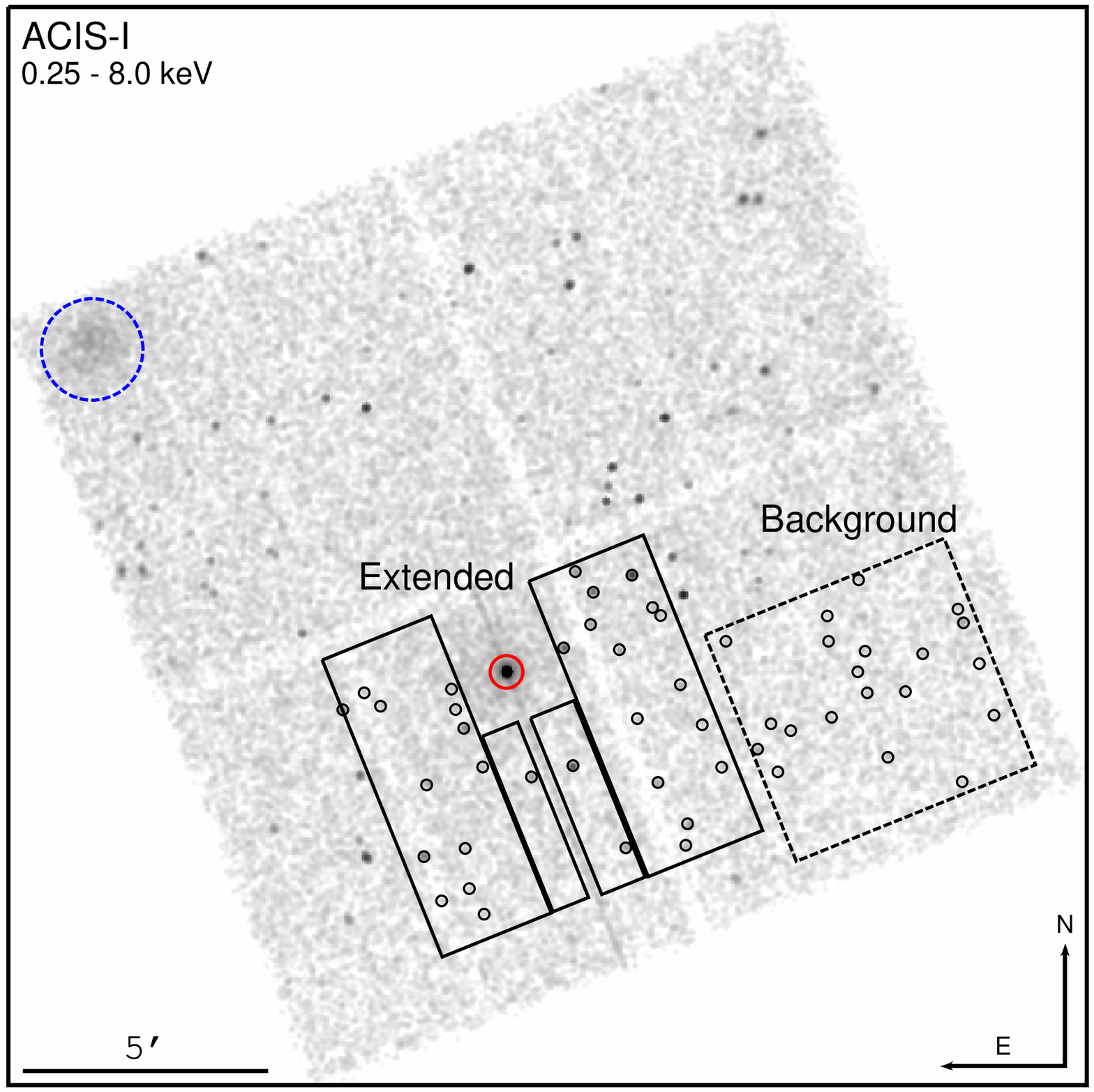}~
\includegraphics[angle=0,width=0.5\linewidth]{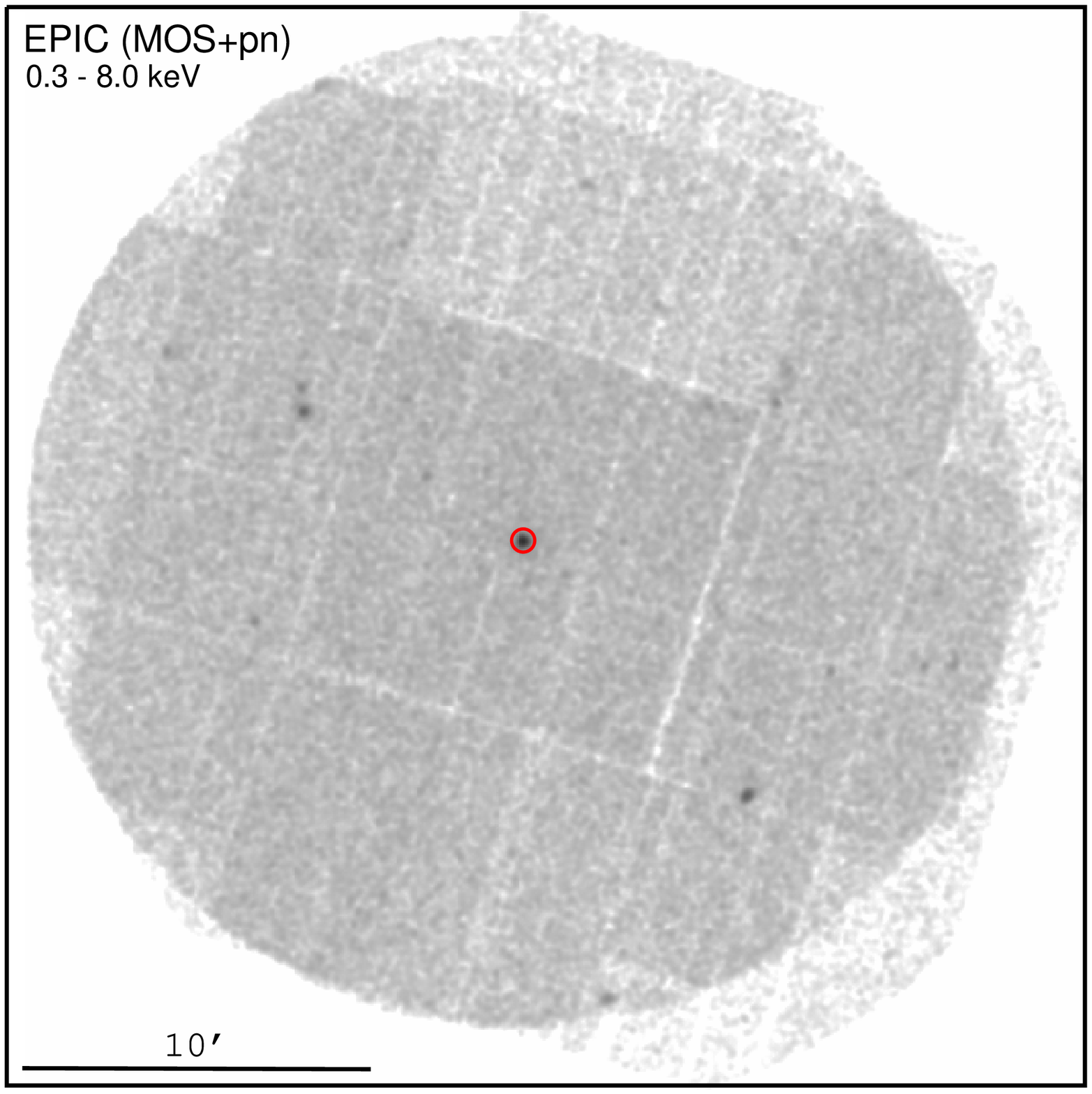}
\caption{FoV of the X-ray observations towards $\zeta$\,Oph and
  BD$+$43$^{\circ}$3654. {\it Left}: {\it Chandra} ACIS-I smoothed
  exposure-corrected image of $\zeta$\,Oph in the 0.25-8.0~keV energy
  range. {\it Right}: {\it XMM-Newton} EPIC (MOS+pn) smoothed
  exposure-corrected image of BD$+$43$^{\circ}$3654 in the
  0.3--8.0~keV energy range. The (red) solid-line circle on each panel
  indicates the spectrum extraction region of the target stars. Other
  point-like sources in the FoV have been identified. The (blue)
  dashed-line circular aperture in the left panel shows the position
  of the diffuse extragalactic source 1AXG\,J163740$-$1027 (see
  text).}
\end{center}
\label{fig:FOV}
\end{figure*}

\begin{figure*}
\begin{center}
\includegraphics[angle=0,width=0.52\linewidth]{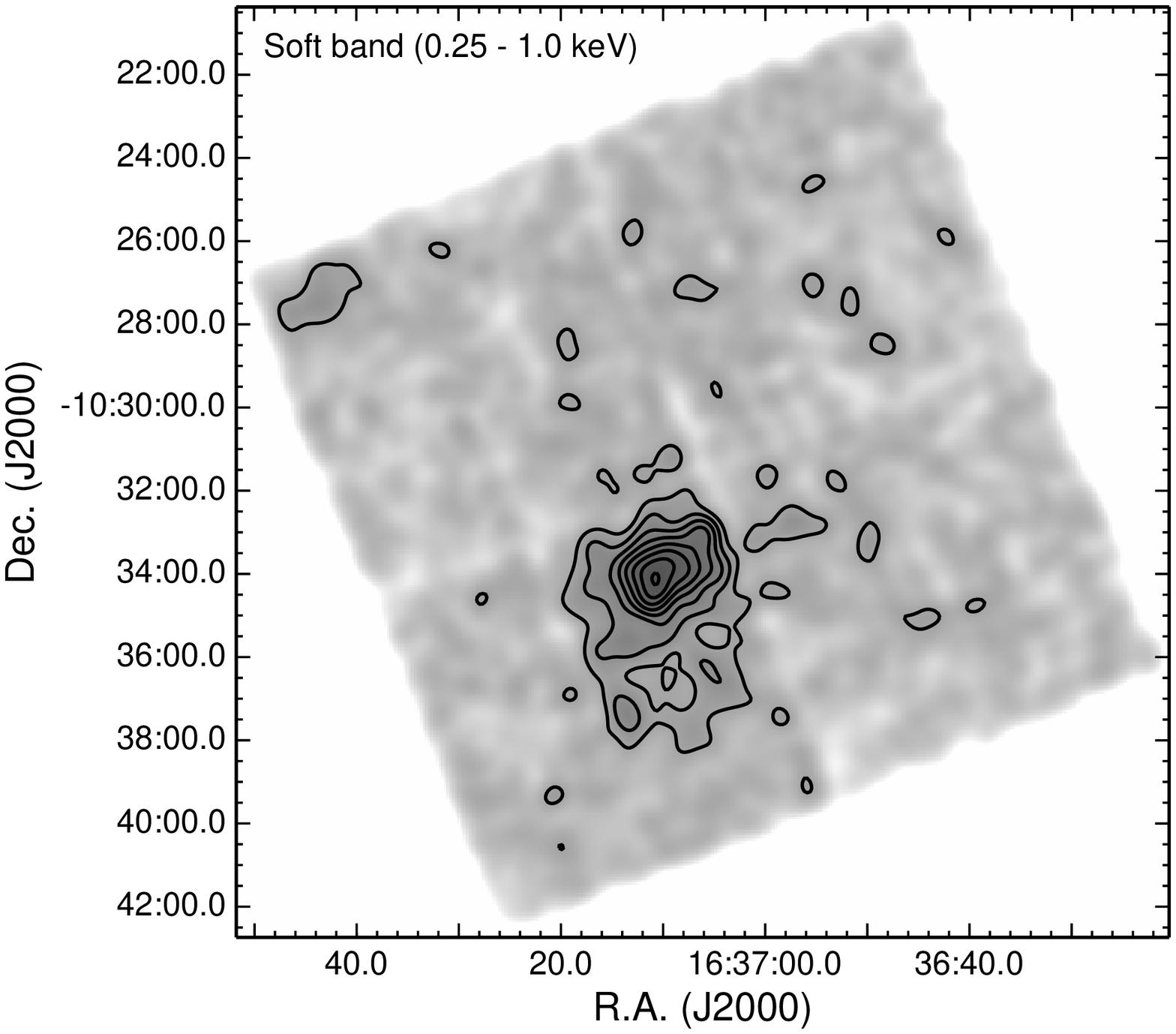}~
\includegraphics[angle=0,width=0.52\linewidth]{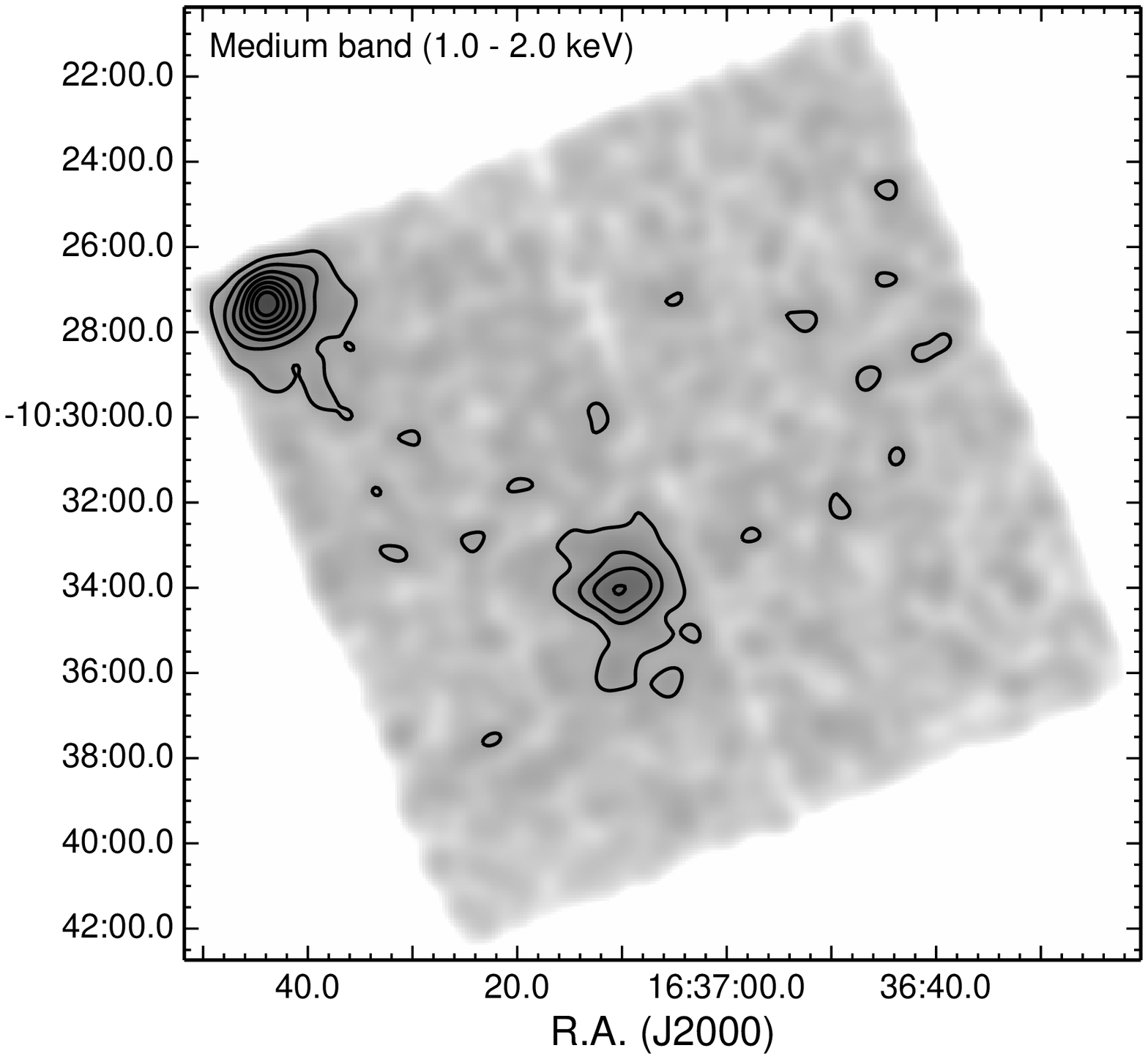}\\
\includegraphics[angle=0,width=0.52\linewidth]{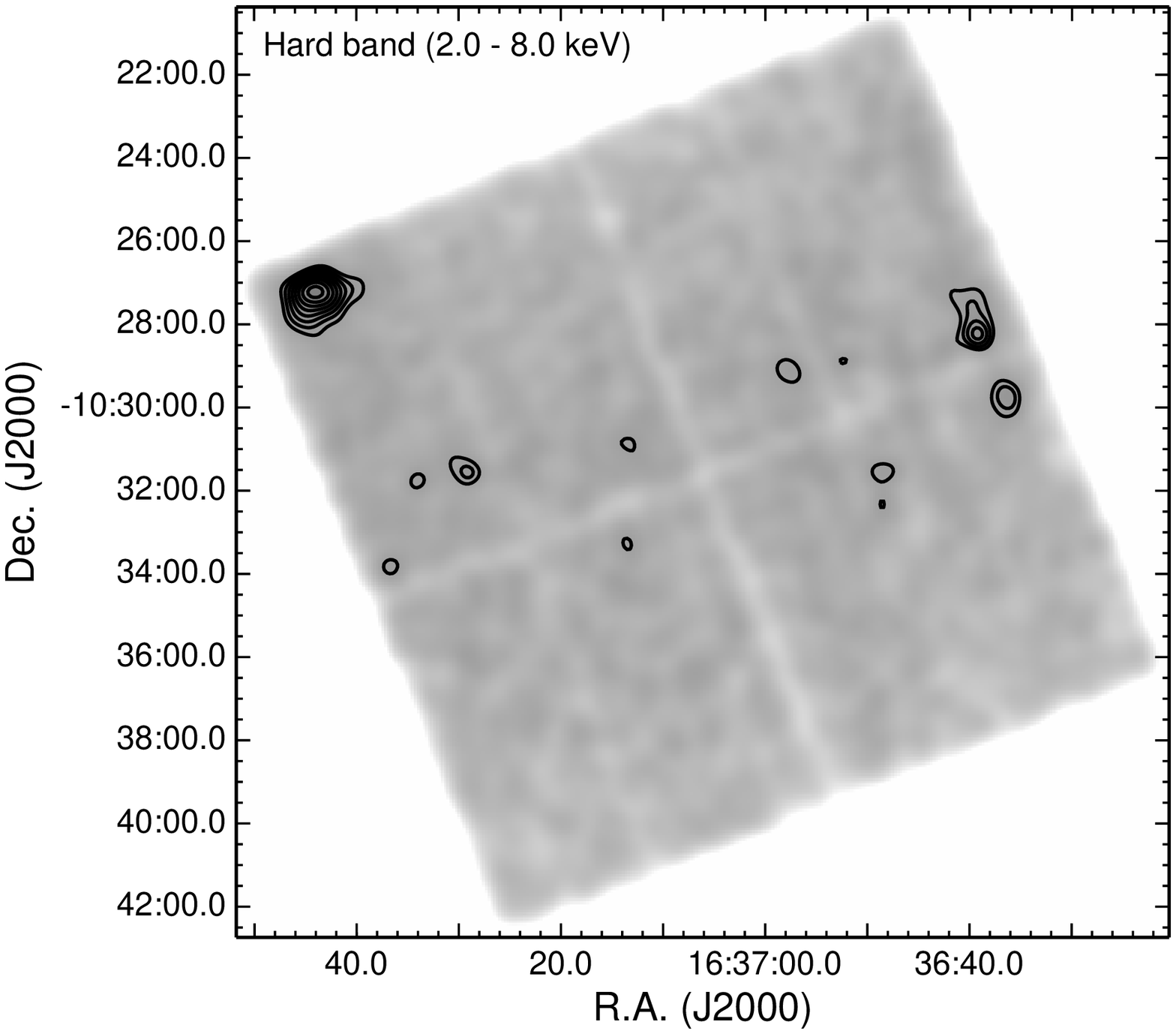}~
\includegraphics[angle=0,width=0.52\linewidth]{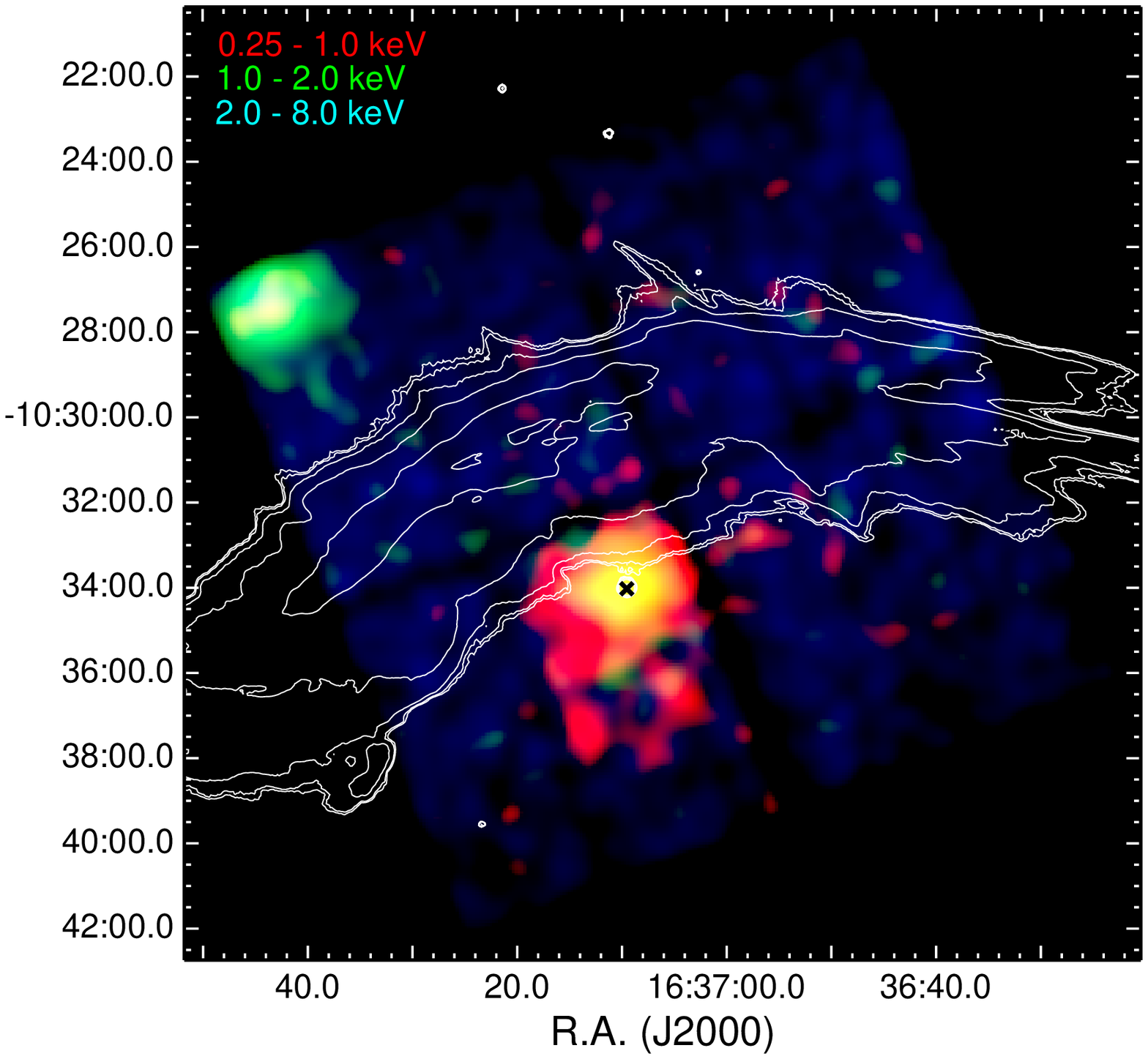}
\caption{{\it Chandra} ACIS-I exposure-corrected,
  background-subtracted images of the X-ray emission around
  $\zeta$\,Oph. The energy bands are labeled on each panel. The
  bottom-right panel shows a color-composite image of the three other
  panels, while contours show the {\it Spitzer} MIPS 24~$\mu$m emission
  from the bow shock. Point sources have been excised from these
  images, including the central star.}
\end{center}
\label{fig:Z_oph_cont}
\end{figure*}

\begin{figure*}
\begin{center}
\includegraphics[angle=0,width=0.52\linewidth]{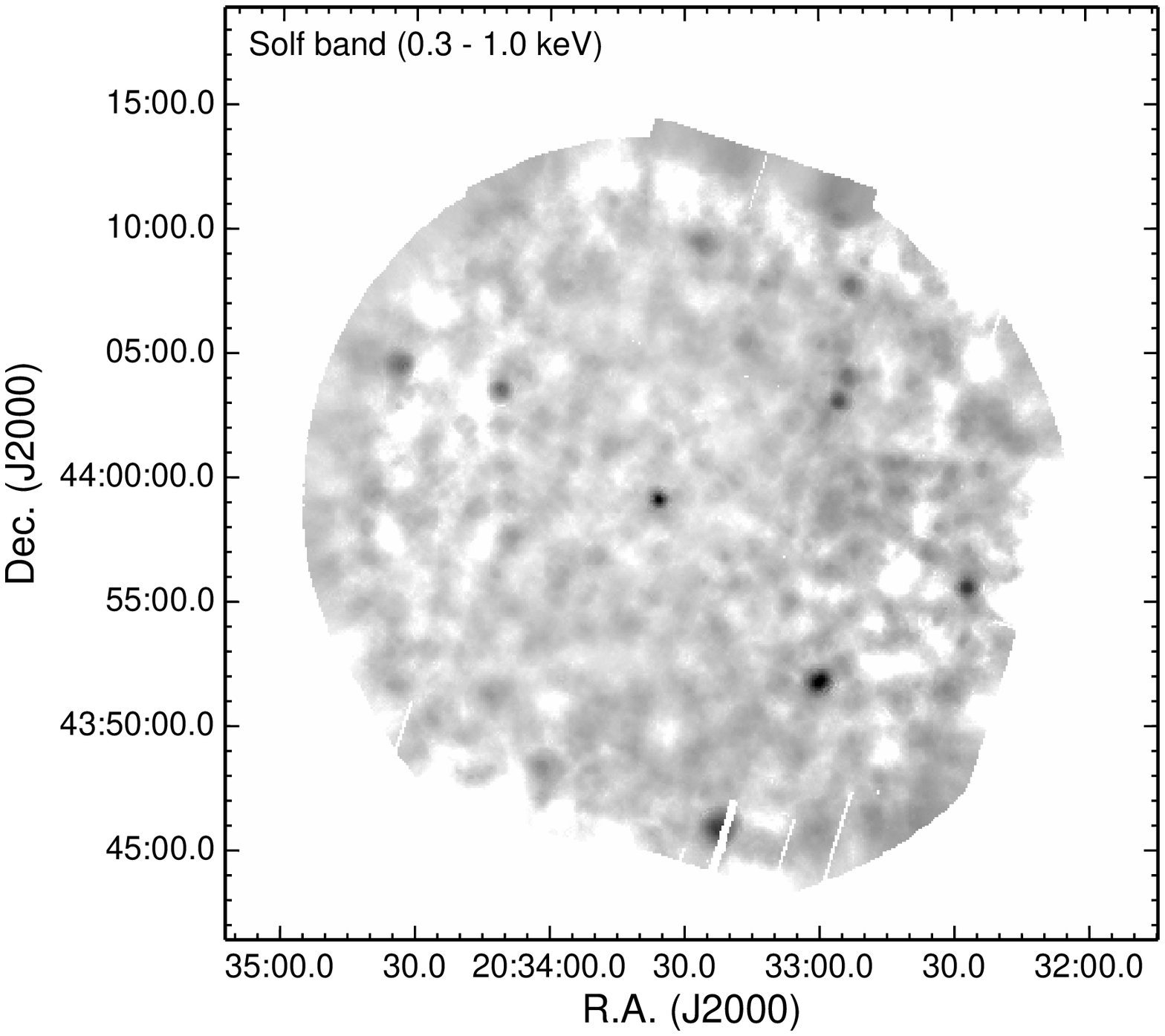}~
\includegraphics[angle=0,width=0.52\linewidth]{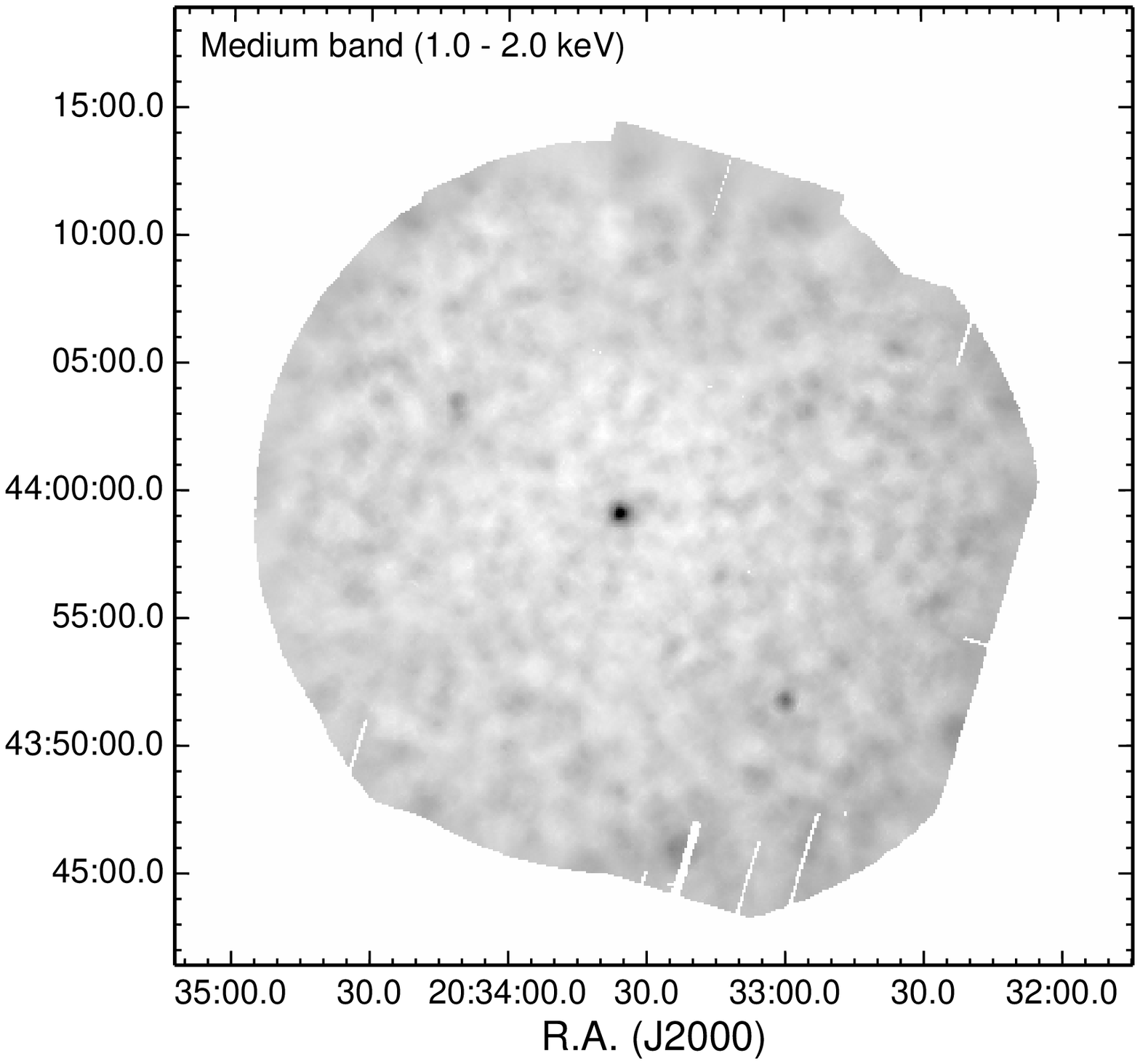}\\
\includegraphics[angle=0,width=0.52\linewidth]{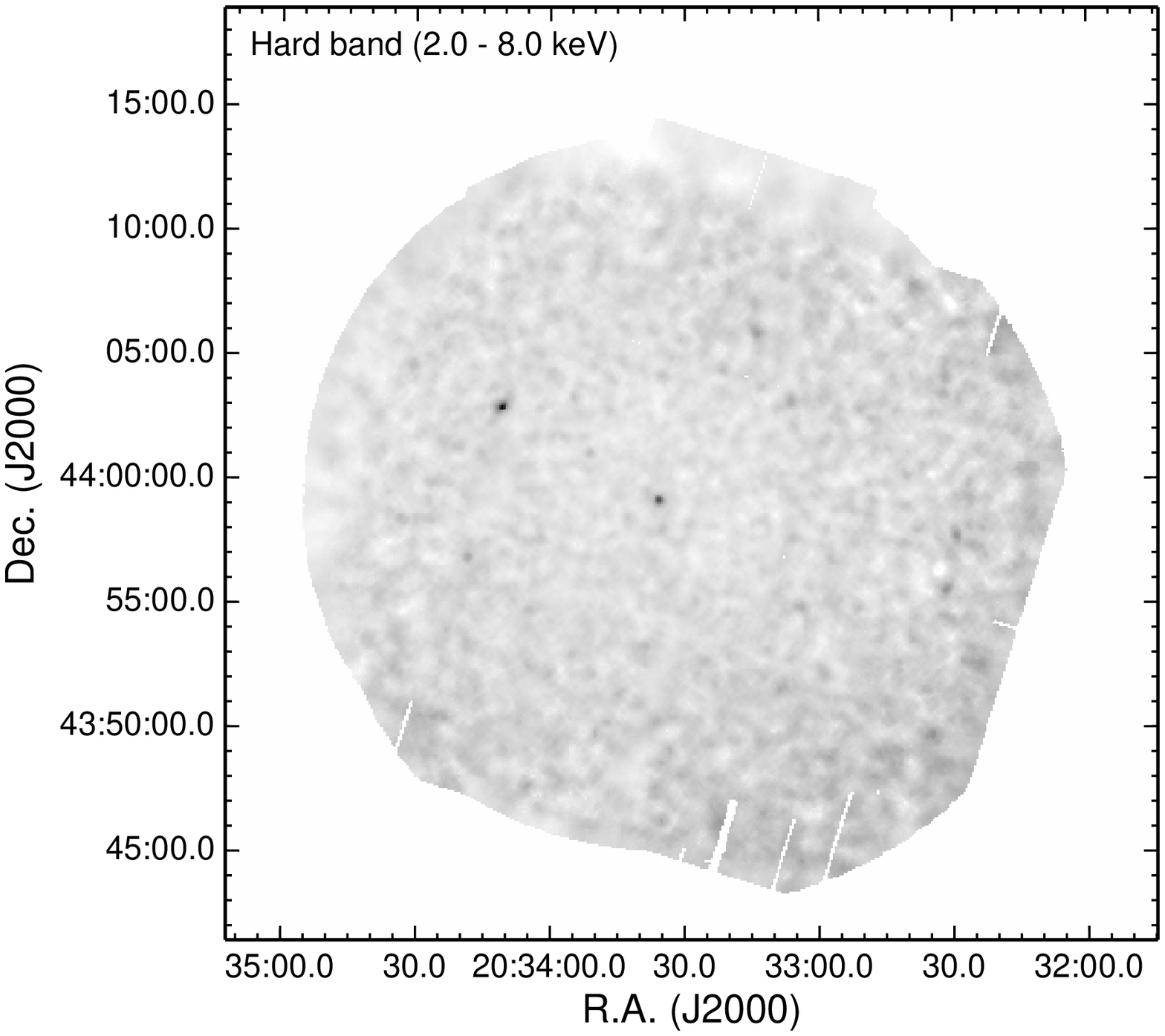}~
\includegraphics[angle=0,width=0.52\linewidth]{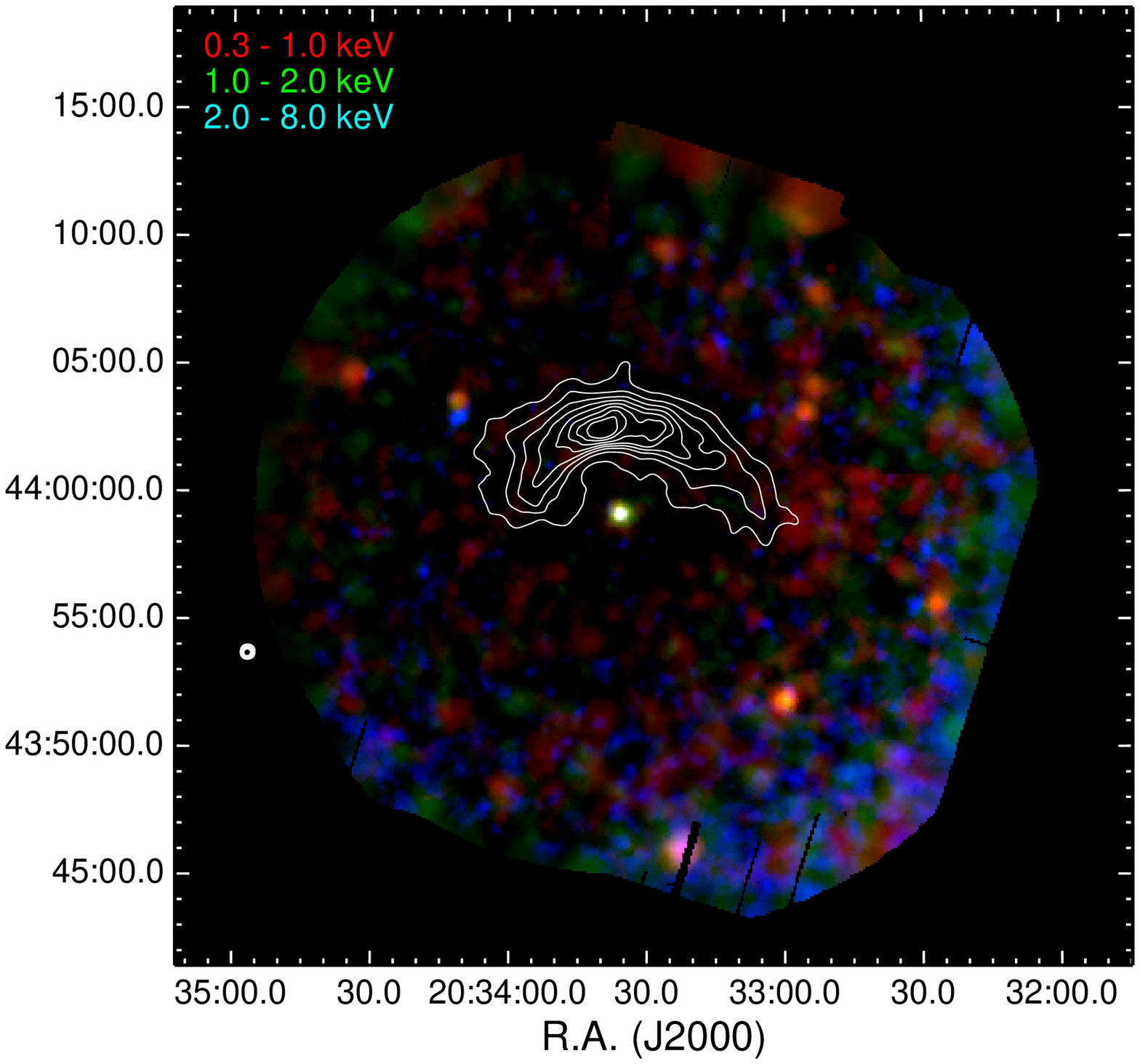}
\caption{{\it XMM-Newton} EPIC (MOS$+$pn) exposure-corrected,
  background-subtracted images of the X-ray emission around
  BD$+$43$^{\circ}$3654. The energy bands are labeled on each
  panel. The bottom-right panel shows a color-composite image of the
  three other panels, while contours show the {\it WISE} 22~$\mu$m
  emission from the bow shock. The star is centered on each
  panel. Unlike Fig.~3, no point sources have been excised from these
  images.}
\end{center}
\label{fig:BD_cont}
\end{figure*}

\subsection{$\zeta$\,Oph} 

The {\it Chandra} observations of $\zeta$\,Oph were performed on 2013
July 3 (Observation ID: 14540; PI: L.M.\,Oskinova) using the Advanced
CCD Imaging Spectrometer (ACIS-I) for a total exposure time of
72.1~ks. The {\it Chandra} Interactive Analysis of Observations (CIAO)
software package version 4.6 \citep{Fruscione2006} was used to analyze
the data using CALB version 4.6.3. The resulting exposure time after
excising dead time periods is 71.8~ks. Figure~2-left panel presents
the field of view (FoV) of the ACIS-I observations in the
0.25--8.0~keV energy range. Several point-like sources can be
identified as well as a diffuse source towards the northeast of the
FoV of the ACIS-I detectors with its maximum located at
(R.A.,Dec.)=(16$^\mathrm{h}$ 37$^\mathrm{m}$
44.2$^\mathrm{s}$,$-$10$^\circ$ 27$\arcmin$ 17.1$\arcsec$). This
source is spatially coincident with 1AXG\,J163740$-$1027 as
reported in the {\it ASCA} Medium Sensitivity Survey by
\citet{Ueda2001} within the error reported by those authors.

Exposure-corrected, background-subtracted images of the soft
(0.25--1.0~keV), medium (1.0--2.0~keV), and hard (2.0--8.0~keV) X-ray
images are presented in Figure~3. Point-like sources have been removed
and the gaps have been filled with the CIAO task {\it dmfilth}. The
final images were smoothed with the CIAO task {\it aconvolve}, with a
Gaussian kernel of 4$\arcsec$ in the brightest regions. A composite
color picture of the three images is presented in Fig.~3-bottom right
panel. White contours show the distribution of the MIPS 24~$\mu$m
emission around $\zeta$\,Oph.

We also used {\it Suzaku} observations of $\zeta$\,Oph to complement
our study. These observations were performed on 2008 March 15
(Observation ID: 402038010; PI: W.L.\,Waldron) using the X-ray Imaging
Spectrometers XIS\,0, XIS\,1, and XIS\,3. The net exposure times for
each camera are 95.7~ks. Due to their lower angular resultion, no
spatial distribution of the X-ray-emitting gas can be performed from
these cameras. We only used the {\it Suzaku} observations to performed
the spectral study of $\zeta$\,Oph (see Section~3.1). To illustrate
this, we show in the Appendix~A the smoothed exposure-corrected image
of the Suzaku XIS\,1.

\subsection{BD+43$^{\circ}$3654} 

The {\it XMM-Newton} observations towards BD+43$^{\circ}$3654 were
performed in 2010 May 8 (Observation ID: 0653690101; PI:
V.\,Zabalza). The EPIC cameras were operated in the full-frame mode
with the thin optical filter for a total exposure time of 38.5, 45.7,
45.7~ks for the EPIC-pn, MOS1, and MOS2 cameras, respectively. The
observations were processed using the {\it XMM-Newton} Science
Analysis Software (SAS version 13.5.0) with the associated calibration
files (CCF) available on 2014 October 28. Figure~2-right panel shows
the FoV of the EPIC observations in the 0.3--8.0~keV energy
range. Unfortunately, the observations were severely affected by
high-background levels and the final net exposure times are 7.0, 23.5,
and 26.6~ks for the EPIC-pn, MOS1, and MOS2 cameras, respectively. We
want to note that \citet{Terada2012} have used these {\it XMM-Newton}
observations to search for point-like sources in the FoV of their {\it
  Suzaku} observations of BD+43$^{\circ}$3654, however, they did not
performed further analysis of the {\it XMM-Newton} data.

Exposure-corrected, background-subtracted images at different bands
(namely soft 0.3--1.0~keV, medium 1.0--2.0~keV, and hard 2.0--8.0~keV
bands) were generated using the ESAS-XMM tasks. The final images have
been adaptively smoothed using the ESAS-XMM task {\it adapt}
requesting 50~counts for the three bands (see Figure~4). A composite
color picture of the three images is presented in Fig.~4-bottom right
panel. White contours show the distribution of the WISE 22~$\mu$m
emission around BD+43$^{\circ}$3654.

\section{Results}
\label{sec:results} 

As expected, both central stars are detected in X-rays. Figures~3 and
4 show the spatial distribution of the X-ray emission around our
targets. As can be seen in Fig.~3 {\it Chandra} images show that
diffuse X-ray emission is present close to $\zeta$\,Oph spatially
coinciding with the likely location of the bow shock wake. On the
other hand, we find no extended X-ray emission associated with the
bow-shock apex.

Figure 4 corroborates the findings presented by \citet{Terada2012},
who did not find any hint of diffuse X-ray emission associated with
the bow shock around BD$+$43$^{\circ}$3654 in their {\it Suzaku}
observations.

\begin{figure*}
\begin{center}
\includegraphics[angle=0,height=1\linewidth]{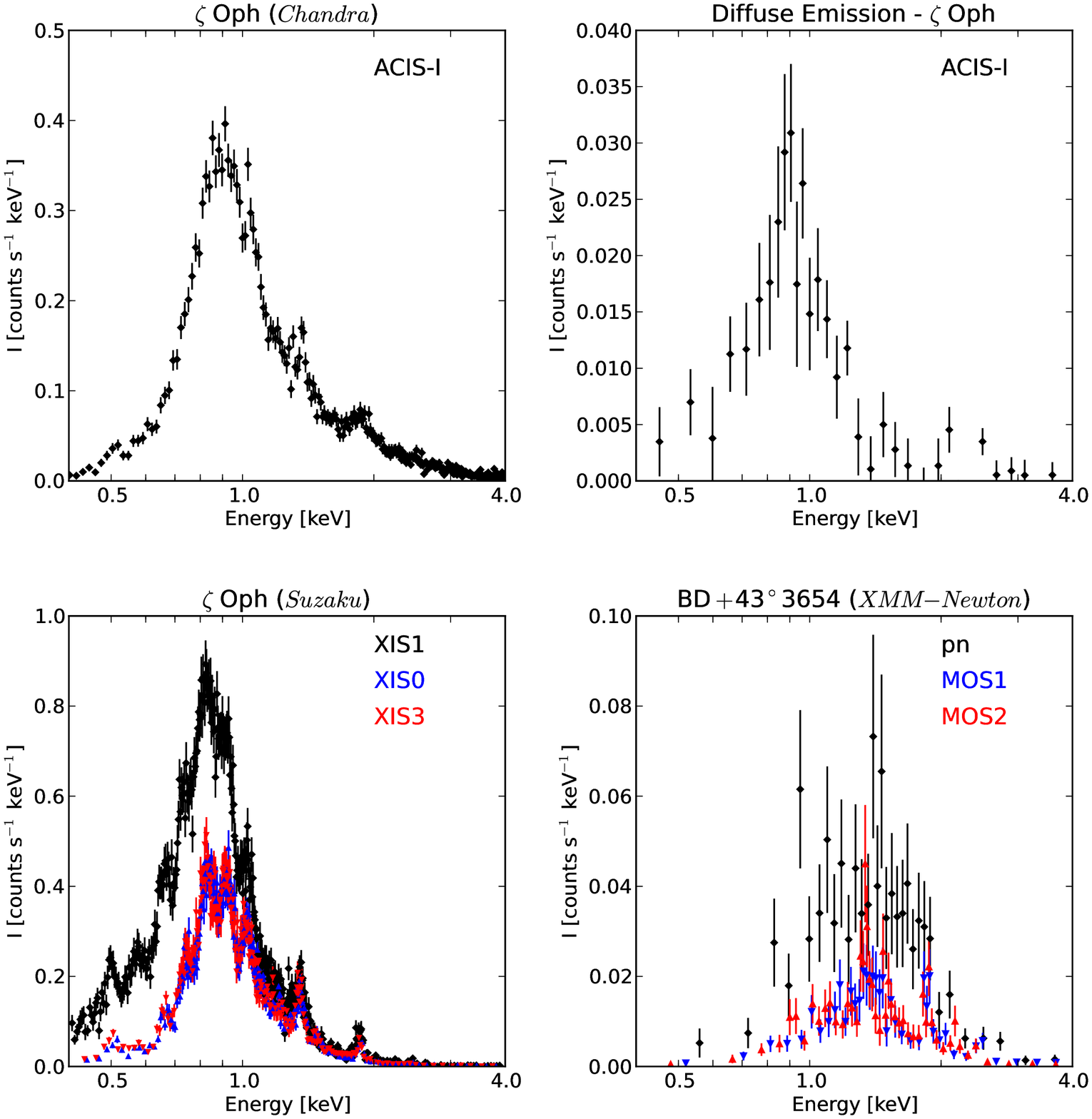}~
\caption{Background-subtracted spectra from different observations
  used in the present paper. {\it Chandra} ACIS-I spectra of
  $\zeta$\,Oph (upper left panel) and its corresponding extended
  emission (upper right panel). The bottom left panel shows the
  spectra of $\zeta$\,Oph obtained by the {\it Suzaku} cameras. The
  bottom right panel shows the spectra of BD$+$43$^{\circ}$3654 as
  obtained by the {\it XMM-Newton} EPIC cameras.}
\end{center}
\label{fig:spec_all}
\end{figure*}

\subsection{X-rays from $\zeta$\,Oph}

We carefully examined the {\it Chandra} images of $\zeta$\,Oph for the
different energies and in full band to search for traces of diffuse
emission associated with the bow shock. However, no such emission was
detected. Hence, if any X-ray emission directly associated to the bow
shock should exist, it shall be below the background level (see
Section~4).

In order to study the physical properties of the X-ray emission from
$\zeta$\,Oph and the apparent extended emission around it, we have
extracted two spectra from the {\it Chandra} ACIS-I observations. A
circular aperture with radius of 20$\arcsec$ has been used to extract
a spectrum from $\zeta$\,Oph, while the corresponding spectrum from
the diffuse emission has been extracted from the polygonal apertures
shown in Fig.~2-left panel. The background region for both spectra has
been selected from a region with no diffuse X-ray emission. The
spectra of $\zeta$\,Oph and its diffuse X-ray emission are presented
in the top panels of Figure~5.

During the analysis of the {\it Chandra} data, we realized that the
spectrum from $\zeta$\,Oph suffered from the effect of
pile-up. Because of this, we decided to analyze the archived {\it
  Suzaku} data of this source. We have extracted the XIS\,0, XIS\,1,
and XIS\,3 spectra from a circular apertures with radii of 4\farcm3
centered at the position of $\zeta$\,Oph, and the background region
has been extracted from an annular region (see Appendix~A for
details). The {\it Suzaku} XIS0, XIS1, and XIS3 spectra are shown in
Figure~5 - bottom left panel.

As expected, the {\it Chandra} and {\it Suzaku} spectra of
$\zeta$\,Oph (Fig.~5 - left panels) present very similar features: a
broad main feature centered at 0.9~keV, with two secondary peaks
around $\lesssim$1.4 and 1.8 keV, and a rapid decay at energies
greater than 3.0~keV. The spectrum of the apparently extended X-ray
emission in vicinity of $\zeta$\,Oph was extracted from the polygonal
regions defined in Fig.~2-left panel excising point-like sources
present in these regions. The spectrum of extended emission is very
similar to the spectrum of the central star (Fig.~5, top right panel),
with maximum of spectral energy distribution at about 0.9~keV and no
significant count rate below 0.4~keV.

To study physical properties of X-rays from $\zeta$\,Oph and the
associated extended X-ray emission, we have performed spectral
analysis using XSPEC \citep[v.12.8.2][]{Arnaud1996}. The fits were
performed taking into account a Tuebingen-Boulder interstellar medium
absorption model as incorporated in XSPEC \citep{Wilms2000}. The
abundances for the star and that of the diffuse emission were assumed
to be the same. We assumed the C, N, and O abundances as those
reported by \citet{Villamariz2005}. The interstellar column density
was fixed according to known reddening of $\zeta$\,Oph at
$N_\mathrm{H}=6\times10^{20}$~cm$^{-2}$ \citep[e.g.][]{Liszt2009}.

We started our modeling of $\zeta$\,Oph using a simple model and then
increased its complexity. We have fit a i) single {\it apec} plasma
temperature, ii) a two-temperature {\it apec} plasma model, iii) a
power law model, iv) a one-temperature {\it apec} plasma model plus a
power law component, and iv) a two-power law model. None of these
combinations could fit the observed spectrum and resulted in fits with
reduced $\chi^{2}$ bigger than 5, unless we include the effects of
pile-up. Apparently the broad prominent spectral feature seen at
around 1.8~keV is just a pile-up effect leading to the doubling of
energies of photons at the maximum of spectral energy distribution
around 0.9~keV. This is likely the reason that the 1.8~keV feature in
the {\it Suzaku} spectra seems narrower, that is, is not affected by
pile-up.

The best-fit model, taking into account the pile-up, resulted in a
$\chi^{2}/\mathrm{d.o.f.}$=1.98 and accounts for the contribution of a
thermal component (an {\it apec} plasma model) and a power law model
(see Table~1). The plasma temperature is
$kT$=0.80$^{+0.02}_{-0.02}$~keV and the power law index of
$\Gamma$=3.05$^{+0.10}_{-0.11}$. Surprisingly, models including one
plasma temperature ($apec$) or two-plasma temperature components
($apec+apec$) did not resulted in a good fit ($\chi^{2}>$5), thus, we
do not listed it in Table~1.

The absorbed and unabsorbed fluxes in the 0.4--4.0 keV energy range
are $f$=2.10$\times$10$^{-12}$~erg~cm$^{-2}$~s$^{-1}$ and
$F$=2.50$\times$10$^{-12}$~erg~cm$^{-2}$~s$^{-1}$, respectively. The
total X-ray luminosity at a distance of 222~pc
\citep[see][]{Megier2009} is
$L_\mathrm{X}$=1.5$\times$10$^{31}$~erg~s$^{-1}$.

Assuming that the extended X-ray emission is a combination of the
spillover created by the pile-up and the diffuse X-ray emission we
have used the best-fit model parameters of $\zeta$\,Oph as components,
plus another component. We note that we used the same ratio of the
normalization parameters from the central star ($A_{1}/A_{2}$=0.93).
We found that the best-fit model was achieved accounting for a two
temperatures components ($apec_{1} + apec_{2}$) and a fixed power law
as obtained from $\zeta$\,Oph model ($\Gamma=3.05$; see
Table~1). The plasma temperature of the diffuse X-ray
emission was found to be $kT$=0.20$^{+0.09}_{-0.07}$~keV
($T_\mathrm{X} = 2.3 \times$10$^{6}$~K). The absorbed and unabsorbed
fluxes of this component resulted to be
$f_\mathrm{DIFF}$=8.4$\times$10$^{-14}$~erg~cm$^{-2}$~s$^{-1}$ and
$F_\mathrm{DIFF}$=1.30$\times$10$^{-13}$~erg~cm$^{-2}$~s$^{-1}$. Its
corresponding luminosity at a distance of 222~pc is
$L_\mathrm{DIFF}$=7.60$\times$10$^{29}$~erg~s$^{-1}$.

Finally, in order to assess the validity of the {\it Chandra} ACIS-I
spectral fits, we have also modeled the X-ray emission as detected by
{\it Suzaku} (Fig.~5 - bottom left panel). These observations do not
have the resolution to spatially separate the X-ray emission from
$\zeta$\,Oph and that of the extended emission. Thus, the {\it Suzaku}
XIS0, XIS1, and XIS3 spectra include both the contribution of the
central star and the putative diffuse emission.

First, we have modelled the X-ray emission as detected by the
back-illuminated CCD XIS\,1. The best fit model resulted in two $apec$
components of $kT_{1}=0.21^{+0.01}_{-0.01}$~keV ($T =
2.4\times$10$^{6}$~K) and $kT_{2}=0.75^{+0.01}_{-0.01}$~keV (see
Table~1). We then performed a joint fit to the three
XIS cameras (XIS\,0$+$XIS\,1$+$XIS\,3) and the best-fit model resulted
in similar parameters (see Table~1). Thus, the {\it Suzaku}
observations also point out at the existence of thermal plasma at
$\sim 2 \times$10$^{6}$~K gas whilst the second component, with plasma
temperature of $kT \approx 0.80$~keV, corresponds to $\zeta$\,Oph.

\subsection{X-rays from BD$+$43$^{\circ}$3654}

We do not detect any hint of diffuse X-ray emission associated to
BD$+$43$^{\circ}$3654 (see Figs. 2 and 4). Not at the position of the
bow shock as in the case of the non-thermal radio emission
\citep[e.g.,][]{Benaglia2010} nor at the position of the wake as in
the case of $\zeta$\,Oph.

In a similar way as in the previous section, we extracted pn, MOS1,
and MOS2 spectra from a circular region with radius of 20$\arcsec$ for
the case of BD$+$43$^{\circ}$3654. The background has been extracted
from a region with no contribution of point sources towards the
south. The resultant background-subtracted EPIC (pn, MOS1, and MOS2)
spectra are shown in Fig.~5-bottom right panel. The EPIC-pn spectrum
exhibit a broad feature around 1.0-2.0~keV, but the MOS spectra
present clearer emission features at 1.4~keV and 1.8~keV. No
significant emission is detected below 0.4~keV and above 4.0~keV.

In order to produce the best-fit model of the X-ray emission from
BD$+$43$^{\circ}$3654, we have fitted the three EPIC spectra (pn,
MOS1, and MOS2) simultaneously. We have used a one-temperature {\it
  apec} optically thin plasma model with solar abundances. We let the
column density ($N_\mathrm{H}$) to be a free parameter in the fit as
it unknown. The best-fit model resulted in an absorbing column density
and plasma temperature of $N_\mathrm{H}=(1.54^{+0.08}_{-0.07}) \times
10^{22}$~cm$^{-2}$ and $kT=0.6^{+0.5}_{-0.4}$~keV with a
$\chi^{2}/\mathrm{d.o.f.}=0.948$. Note that \citet{Terada2012} found
very similar values from their analysis of {\it Suzaku} observations.
More sophisticated models, e.g., a two-temperature plasma emission
model or a power law contribution, did not improved the spectral fits,
on the contrary, they resulted in models with
$\chi^{2}/\mathrm{d.o.f.}<0.8$.

The absorbed flux in the 0.4--4.0~keV energy range is
$f$=1.20$\times$10$^{-13}$~erg~cm$^{-2}$~s$^{-1}$ that corresponds to
an intrinsic flux of
$F$=3.15$\times$10$^{-12}$~erg~cm$^{-2}$~s$^{-1}$. The X-ray
luminosity at a distance of 1.4~kpc \citep[see][and references
  therein]{Comeron2007} is
$L_\mathrm{X}$=7.4$\times$10$^{32}$~erg~s$^{-1}$.

\section{Discussion}

So far, high-energy non-thermal emission is eluding detection in bow
shocks around massive runaway O-type stars \citep[e.g.][and this
  work]{Terada2012,Schulz2014}. We would have expected that, if
present, non-thermal X-ray emission should be spatially coincident
with the bow shock detected in mid-IR wavelengths, but this is not the
case for the two objects studied in the present paper. In particular,
the lack of non-thermal diffuse X-ray emission from
BD$+$43$^{\circ}$3654 is puzzling as VLA observations assured the
nature and presence of non-thermal particles. Although one might argue
that the current {\it XMM-Newton} observations are not sensitive
enough, \citet{Terada2012} did not find neither any signature of
extended emission with their {\it Suzaku} observations as mentioned
previously.

In order to set an upper limit to the non-thermal X-ray emission we
have extracted the background-subtracted spectra of the two
observations from regions spatially coincident of that of the bow
shock, that is, where the non-thermal emission is expected. The
corresponding background count rate in the 0.4--4.0~keV energy range
for $\zeta$\,Oph and BD$+$43$^{\circ}$3654 are 1.1$\times$10$^{-3}$
ACIS-I counts~s$^{-1}$ and 3.7$\times$10$^{-3}$ EPIC-pn
counts~s$^{-1}$, respectively. Using the {\it Chandra} PIMMS
tool\footnote{\url{http://cxc.harvard.edu/toolkit/pimms.jsp}} we can
estimate upper limits to the fluxes and luminosities. If we assume
that the background emission can be modelled by a power law spectrum
with $\Gamma=1.5$, the estimated upper limits to the absorbed
(unabsorbed) fluxes for $\zeta$\,Oph and BD$+$43$^{\circ}$3654 are
9.5\,(10.6)$\times10^{-15}$~erg~cm$^{-2}$~s$^{-1}$ and
1.5\,(3.6)$\times10^{-14}$~erg~cm$^{-2}$~s$^{-1}$, whilst their
corresponding normalization parameters are
2.4$\times$10$^{-6}$~cm$^{-5}$ and 8.2$\times$10$^{-6}$~cm$^{-5}$,
respectively. 
The estimated upper limit to the X-ray luminosity in the 0.4--4.0~keV
energy range is 6.2$\times$10$^{28}$~erg~s$^{-1}$ and
8.4$\times$10$^{30}$~erg~s$^{-1}$ for $\zeta$\,Oph and
BD$+$43$^{\circ}$3654, respectively. Note that \citet{Terada2012}
estimated an upper X-ray luminosity of
$1.1\times$10$^{32}$~erg~s$^{-1}$ for the 0.5--10~keV energy range for
their {\it Suzaku} observations of BD$+$43$^{\circ}$3654 for a photon
index $\Gamma$=1.1. If non-thermal X-ray emission is produced, as
suggested by analytical predictions \citep[e.g.,][]{delValle2012}, its
intensity should be below the background detection levels of the
current X-ray satellites.

Extreme care should be taken when considering the {\it Chandra}
observations of $\zeta$\,Oph, as they have been affected by
pile-up. It must be noted that due to this effect, the final best-fit
model of $\zeta$\,Oph ($apec+$power law) is not to be taken as
definite physical parameters of the star (specifically the power law
component). This model should only be taken as the statistically
best-fit model within the instrumental limitations. Anyhow, it helped
us to restrict the physical origin of the extended emission, a thermal
nature, in addition with the analysis of the {\it Suzaku} data. 

The soft plasma temperature of this extended X-ray emission
($T_\mathrm{X}\approx2\times10^{6}$~K) implies the existence of a
mixing region between the adiabatically-shocked wind region
($T$=10$^{7}$-10$^{8}$~K) and the ionized outer material
($T\approx$10$^{4}$~K), similar to that found in classic wind-blown
bubbles \citep[e.g., H\,{\sc ii} regions, planetary nebulae,
  Wolf-Rayet nebulae, and
  superbubbles;][]{Chu2001,Gudel2008,Jaskot2011,Toala2015}. Simulations
presented by \citet{Mackey2015} suggest that, in the case of runaway
stars, the most important mixing region is placed at the wake of the
bow shock, which would produce a cometary-like distribution of
X-ray-emitting gas. If this is the case for $\zeta$\,Oph, it would be
the first wind-blown bubble around a single O-type star with diffuse
X-ray emission.

We have also examined the archived {\it Chandra} HETG observations of
$\zeta$\,Oph (Obs.\,ID:3857 and 2571) and found no evidence of this
extended X-ray emission in the zero order images.

Finally, it is interesting to discuss the absence of thermal X-ray
emission at the wake of BD$+$43$^{\circ}$3654. Even though this star
has a greater mechanical wind luminosity than $\zeta$\,Oph and can
easily carve an adiabatically-shocked hot bubble due to its high
stellar wind velocity ($v_{\infty} \approx$ 2250~km~s$^{-1}$; see
Section~1), it does not exhibit diffuse X-ray emission. This might be
due to the fact that the wake region in BD$+$43$^{\circ}$3654 seems to
be more contaminated by ISM material in the line of sight than
$\zeta$\,Oph (as illustrated in Fig.~1). Moreover, this region is
detected at the edge of the EPIC cameras which have a reduce
sensitiviy as compared to the central regions.

\section{Conclusions}

We have presented {\it Chandra}, {\it Suzaku}, and {\it XMM-Newton}
observations of the runaway O-type stars $\zeta$~Oph and
BD$+$43$^{\circ}$3654 to investigate the presence of diffuse
non-thermal X-ray emission associated to their bow shocks. We found no
evidence of such X-ray emission associated to the bow
shocks. Nevertheless, we have estimated upper limits for the
non-thermal X-ray luminosity in the 0.4--4.0~keV energy range of
6.2$\times$10$^{28}$~erg~s$^{-1}$ and
8.4$\times$10$^{30}$~erg~s$^{-1}$ for $\zeta$\,Oph and
BD$+$43$^{\circ}$3654, respectively.

Although our {\it Chandra} observations of $\zeta$\,Oph suffered from
pile-up, we are able to detect diffuse thermal emission with plasma
temperature of $T_\mathrm{X} \approx 2\times10^{6}$~K. The
distribution and location of this diffuse X-ray emission in the wake
of the bow shock provides observational support to the predictions of
radiation-hydrodynamic results by \citet{Mackey2015}. This makes
$\zeta$\,Oph the first wind-blown bubble around a single O-type star
that exhibits diffuse X-ray emission.

Future deep {\it XMM-Newton} observations of the present sources could
help improve our findings and put new observational constrains to the
current growing body of theoretical models.

\section*{Acknowledgments}

We would like to thank the anonymous referee for helpful comments that
helped improve our paper. JAT and MAG are supported by the Spanish
MICINN (Ministerio de Ciencia e Innnovaci\'{o}n) grant AYA
2014-57280-P. JAT and LMO acknowledge support from the ISM-SPP DFG
Priority Program 1573. RI expresses appreciation for support that was
provided by the National Aeronautics and Space Administration through
Chandra Award Number G03-14008X, issued by the {\it Chandra} X-ray
Observatory Center, which is operated by the Smithsonian Astrophysical
Observatory for and on behalf of the National Aeronautics Space
Administration under contract NAS8-03060.

\begin{appendix}

\section{Suzaku observations}
\label{app:appendixA}

Figure~A1 presents an image of the {\it Suzaku} XIS1 event file. The
extraction region corresponds to a circular aperture of 4\farcm3 in
radius centered at the position of $\zeta$\,Oph. The background was
extracted from an annular region with inner and outer radii of
4\farcm43 and 7$\arcmin$, respectively.

\begin{figure}
\begin{center}
\includegraphics[angle=0,width=1\linewidth]{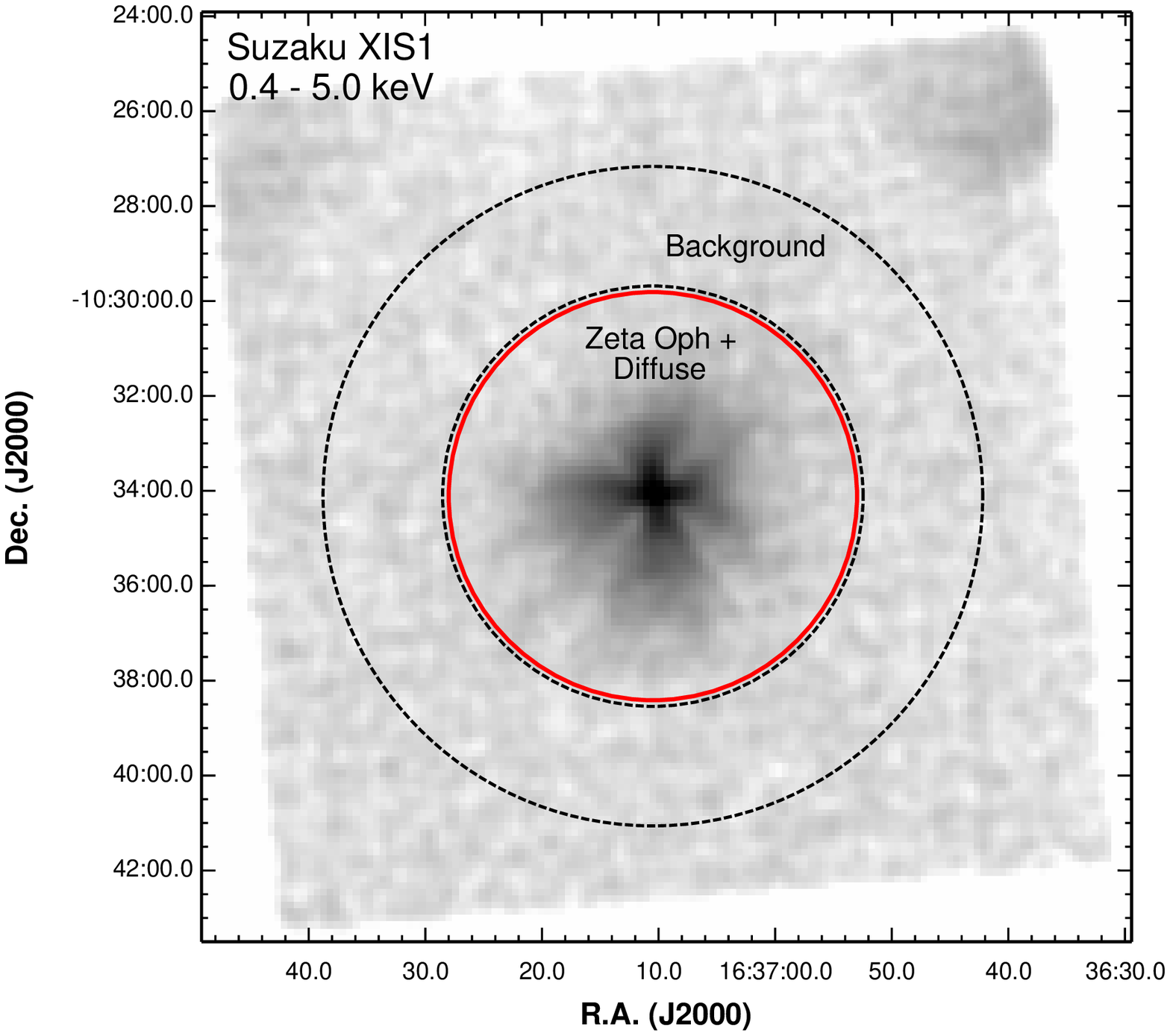}
\caption{{\it Suzaku} XIS1 smoothed image of $\zeta$\,Oph in the
  0.4--5.0~keV energy range. The (red) solid-line circle has a radius
  of 4.3$\arcmin$ and encloses the spectrum extraction region. The
  dashed-line annulus corresponds to the background region.}
\end{center}
\label{fig:figa1}
\end{figure}

\end{appendix}

\begin{table}
\centering
\caption{Best-fit models for spectra obtained from {\it Chandra} and {\it Suzaku} observations}
\begin{tabular}{clclcll}
\hline
\hline
$\zeta$\,Oph$^{\dagger}$ & & & & &                                                     & $\chi^{2}$/DoF \\
   & $apec_{1}$                           &$+$& Power Law                           &   &                   &   1.98=236.23/119 \\ 
   & $kT$=0.80$^{+0.02}_{-0.02}$~keV        &   & $\Gamma=3.05^{+0.10}_{-0.11}$         &   &                   &            \\
   & $A_{1}=$5.2$\times$10$^{-4}$~cm$^{-5}$&   & $A_{2}$=5.6$\times$10$^{-4}$~cm$^{-5}$&   &                   & 	         \\
   &                                      &   & 				    &   & 		    &		 \\
   &                              & &                              &    &                    &  \\ 
\hline
Diffuse Emission & & & & & & \\
around $\zeta$\,Oph & & & & & & \\
   & $apec_{1}$                           &$+$& Power Law                             &   &              & 1.24=38.52/31\\ 
   & $kT_1$=0.75$^{+0.10}_{-0.11}$~keV      &   & $\Gamma=3.05$ (fixed)                 &   &              &             \\
   & $A_{1}=$3.4$\times$10$^{-5}$~cm$^{-5}$ &   & $A_{2}=1.9\times10^{-5}$~cm$^{-5}$     &   &              & 	         \\
   &                                      &   & 				      &   & 		 &		 \\
   & $apec_{1}$                           &$+$& Power Law                             &$+$& $apec_{2}$                       & 1.05=30.58/29\\ 
   & $kT_1$=0.85$^{+0.20}_{-0.19}$~keV      &   & $\Gamma=3.05$ (fixed)                 &   & $kT_2$=0.20$^{+0.09}_{-0.07}$~keV &             \\
   & $A_{1}=$2.0$\times$10$^{-5}$~cm$^{-5}$ &   & $A_{2}=1.1\times10^{-5}$~cm$^{-5}$     &   & $A_{3}=1.2\times10^{-4}$~cm$^{-5}$   & 	         \\
   &                                      &   & 				      &   & 		    &		 \\
\hline
\hline
$\zeta$\,Oph$+$& & & & &                                                                        &       \\
Diffuse Emission& & & & &                                                                        &      \\
XIS1 &  &   &            &   &                   &    \\ 
     & $apec_{1}$                             &$+$& Power Law                            &   &                   &   1.91=2975/1556 \\ 
     & $kT_{1}$=0.65$^{+0.01}_{-0.01}$~keV      &   & $\Gamma=3.8^{+0.5}_{-0.5}$             &   &                   &   \\
     & $A_{1}=$9.5$\times$10$^{-4}$~cm$^{-5}$  &   & $A_{2}$=4.5$\times$10$^{-4}$~cm$^{-5}$ &   &                   &  \\
     &                                        &   & 			                 &   & 		       &  \\
XIS1 &                                        &   &  &    &                                           &            \\ 
     & $apec_{1}$                             &   &  & $+$   & $apec_{2}$                                &   1.52=2371.61/1556 \\ 
     & $kT_{1}$=0.75$^{+0.01}_{-0.01}$~keV      &   &  &       & $kT_{2}=0.21^{+0.01}_{-0.01}$~keV           &      \\
     & $A_{1}=$9.3$\times$10$^{-4}$~cm$^{-5}$  &   &  &       & $A_{3}$=2.1$\times$10$^{-3}$~cm$^{-5}$     &      \\
     &                                       &   &   &       &			                          & 	  \\
\hline
XIS0+XIS1+XIS3 &  &   &            &   &                   &    \\ 
          & $apec_{1}$                & &    &$+$& $apec_{2}$                   &                  1.50=5658.75/3769\\ 
          & $kT_{1}=0.74^{+0.01}_{-0.01}$keV & & &  & $kT_{2}$=0.20$^{+0.01}_{-0.01}$~keV  &          \\
          & $A_{1}$=9.4$\times$10$^{-4}$~cm$^{-5}$  & & &  & $A_{3}=$2.1$\times$10$^{-3}$~cm$^{-5}$  &              \\
          &                              &   & 	& &		           &    		   \\
XIS0+XIS1+XIS3 &  &   &            &   &                   &    \\ 
          & $apec_{1}$                            &$+$&  Power Law                          &$+$&$apec_{2}$     & 1.89=741.5/394\\ 
          & $kT_{1}=0.77^{+0.01}_{-0.01}$~keV       &   &  $\Gamma=3.43^{+0.10}_{-0.10}$        &   &  $kT_{2}$=0.24$^{+0.01}_{-0.01}$~keV      &   \\
          & $A_{1}$=7.2$\times$10$^{-4}$~cm$^{-5}$ &   &  $A_{2}=3.1\times10^{-4}$~cm$^{-5}$   &   & $A_{3}=$1.4$\times$10$^{-3}$~cm$^{-5}$       &  \\
          &                                       &   & 			             &   & 		   &  \\
\hline
\hline
\end{tabular}
\begin{list}{}{}
\item{$^{\dagger}$Model performed accounting for the pile-up effect.}
\end{list}
\label{tab:table1}
\end{table}

\end{document}